\documentclass[aps,pre,twocolumn,superscriptaddress,floatfix,showpacs]{revtex4}
\usepackage{graphicx}
\usepackage{bm}
\bibstyle{apsrev.bib}

\begin{document}
\title{Disorder Induced Phases in Higher Spin Antiferromagnetic 
Heisenberg Chains}
\author{Enrico Carlon}
\altaffiliation{Current address: Interdisciplinary Research Institute c/o IEMN, Cit\'e
Scientifique BP 69, F-59652 Villeneuve d'Ascq, France}
\affiliation{Theoretische Physik, Universit\"at des Saarlandes,
D-66041 Saarbr\"ucken, Germany}
\author{P\'eter Lajk\'o}
\affiliation{Department of Physics, Kuwait University, 
P. O. Box 5969, Safat, Kuwait 13060}
\affiliation{Institute of Theoretical Physics, Szeged University, 
H-6720 Szeged, Hungary}
\author{Heiko Rieger}
\affiliation{Theoretische Physik, Universit\"at des Saarlandes,
D-66041 Saarbr\"ucken, Germany}
\author{Ferenc Igl\'oi}
\affiliation{Institute of Theoretical Physics, Szeged University,
H-6720 Szeged, Hungary}
\affiliation{Research Institute for Solid State Physics and Optics, 
H-1525 Budapest, P.O.Box 49, Hungary}

\date{\today}

\begin{abstract}

Extensive DMRG calculations for spin $S=1/2$ and $S=3/2$ disordered
antiferromagnetic Heisenberg chains show a rather distinct behavior in
the two cases.  While at sufficiently strong disorder both systems are
in a random singlet phase, we show that weak disorder is an irrelevant
perturbation for the $S=3/2$ chain, contrary to what expected from a
naive application of the Harris criterion. The observed irrelevance is
attributed to the presence of a new correlation length due to enhanced
end-to-end correlations. This phenomenon is expected to occur for all
half-integer $S > 1/2$ chains. A possible phase diagram of the chain
for generic $S$ is also discussed.

\end{abstract}


\maketitle

\newcommand{\bc}{\begin{center}}
\newcommand{\ec}{\end{center}}
\newcommand{\be}{\begin{equation}}
\newcommand{\ee}{\end{equation}}
\newcommand{\ba}{\begin{array}}
\newcommand{\ea}{\end{array}}
\newcommand{\beqn}{\begin{eqnarray}}
\newcommand{\eeqn}{\end{eqnarray}}

\section{Introduction}

A simple model where the effects of the interplay between quantum
fluctuations and disorder can be studied in great detail is the spin-$S$
random antiferromagnetic Heisenberg (AFH) chain with Hamiltonian:
\begin{equation}
H = \sum_i J_{i} \vec{S}_i \cdot \vec{S}_{i+1}
\end{equation}
where the $J_i > 0$ are quenched random variables.

In absence of randomness ($J_{i} = J$) quantum fluctuations lead to
qualitatively different behavior for half-integer ($S=1/2, 3/2\ldots$) and
integer ($S=1,2\ldots$) values of the spin \cite{Hald81}.  Half-integer
spin chains have a gapless spectrum and quasi long range order.  It is
believed that they all belong to the same (bulk) universality class
independently on $S$ \cite{s32}. This was explicitly verified numerically
for the $S=3/2$ chain \cite{Hall96}, which was found to have the same
bulk decay exponent as for the $S=1/2$ chain.  Integer spin chains are
instead gapped and have a hidden topological order \cite{denN89}.

The effect of quenched randomness on the AFH chains can be studied
by the Strong Disorder Renormalization Group (SDRG) method.  This
technique consists in successively decimating out the strongest bond
in the chain \cite{mdh}.  For the $S=1/2$ chain the SDRG procedure can
be carried out in great detail and yields a probability distribution
of coupling constants that under renormalization is broadened without
limits and in the so called infinite randomness fixed point the method
becomes asymptotically exact. At this fixed point, describing the
so-called random singlet (RS) phase \cite{fisherxx}, the energy scale
($\Gamma$) and the length scale ($\xi$) are related via $\ln \Gamma \sim
\xi^{\psi}$, which differs from scaling at a conventional (random) fixed
point, where $\Gamma \sim \xi^{-z}$ with a finite dynamical exponent $z$.
Numerical studies on random spin $S=1/2$ AFM chains generally agree with
the SDRG results \cite{henelius,ijr00}, although some issues are still
debated \cite{stolze}.

For $S>1/2$ the SDRG procedure requires a higher degree of approximations
with respect to the $S=1/2$ case. A spin $S$ is represented by a
maximally symmetrized combination of $2S$ identical $S=1/2$ spins. The
SDRG decimation then produces effective spins of magnitudes $S_{\rm
eff} \leq S$ obtained by linking the spin-$1/2$ objects into correlated
singlets. For the $S=1$ case an SDRG analysis shows that a sufficiently
strong randomness induces a RS phase with $S_{\rm eff} = 1$, separated
from the gapped phase by a gapless region of Griffiths singularities
\cite{griffiths} (the gapless Haldane phase \cite{Hyma97}) in which
the dynamical exponent, $z$, increases with disorder.  A recent SDRG
study \cite{Refa02} of the disordered AFH chain with $S=3/2$ indicated
the existence of two phases: At strong disorder the relevant degrees of
freedom are $S_{\rm eff} = 3/2$ spins, while at weaker disorder they are
of type $S_{\rm eff} = 1/2$.  Both phases are of RS type with identical
critical exponents.  A quantum critical point separating them is expected
to have specific multicritical exponents \cite{Refa02,Daml02}.

As the SDRG method might fail at weak disorder, its predictions should
be tested by means of accurate numerical analysis, which is the aim
of the present paper. We will present numerical evidences, supported
by theoretical arguments, that the phase diagram for the $S=3/2$ spin
chain differs from what predicted by SDRG \cite{Refa02}. Our numerical
results support the existence of a RS phase only for sufficiently strong
disorder, while weak disorder appear to be an irrelevant perturbation
for the system.

\section{Numerical results}

Here we analyze the critical behavior of the AFH chains with varying
degree of disorder by means of density matrix renormalization group
(DMRG) techniques \cite{DMRGbook}. Numerical calculations are restricted
to the $S \leq 3/2$ case, and the disorder average is performed on $10^3$
-- $10^4$ samples taken from the distribution:
\be
p_\delta (J) = \delta^{-1} J^{-1+1/\delta}
\ \ \ \ \ \ \ \ \ \
{\rm for} \ 0 \leq J \leq 1
\label{distribut}
\ee
where $\delta^2 ={\rm var}[ \ln J]$ measures the strength of disorder.
As usual in DMRG \cite{DMRGbook} we use open boundary conditions and
keep typically $m=80-100$ states in the renormalization procedure.
In order to obtain numerically stable results we use the DMRG finite
system method performing several "sweeps" ($\sim 4-5$) through the chain
\cite{DMRGbook}, which is an essential step in the DMRG procedure.
The DMRG calculations for the $S=3/2$ chain have been extended up to
$L=32$ for weak disorder, while at stronger disorder the procedure tends
to become less stable and we had to restrict ourselves to shorter chains.
However the physical behavior in the strong disorder regime turns out
to be quite well understood, while, in view of the limitations of the
SDRG at weak disorder, it is precisely this regime which is physically
the most interesting.  In the weak disorder regime the DMRG is a rather
stable and reliable technique.

\begin{figure}[t]
\includegraphics[width=7.8cm]{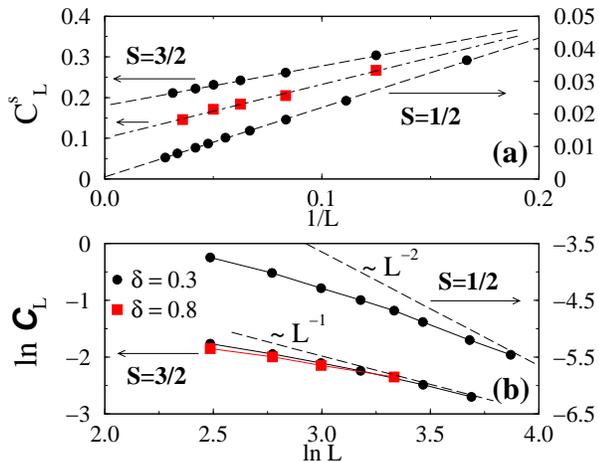}
\caption{Average correlations in the random $S=1/2$ and $3/2$ chains with
disorder strengths $\delta=0.3$ (circles) and $\delta=0.8$ (squares). (a)
End-to-end correlations; dashed lines are linear fits of the data.(b)
${\cal C}_L$ vs $L$ in a log-log scale. Here the dashed line show the
decay in the pure system $(1/L)$ and in a RS phase $1/L^2$.}
\label{FIG01}
\end{figure}

\subsection{Correlation functions}

In order to emphasize the differences between the two cases we
present together the results for the $S=3/2$ and $S=1/2$ chains for
the spin-spin correlation function defined as:
\be
C(i,j)=[\langle S^{z}_i S^{z}_j \rangle]_{\rm av}
\label{cij}
\ee
where $[\dots ]_{\rm av}$ denotes the averaging over quenched disorder.
The correlation function obviously does not depend on the spin direction.

Figure \ref{FIG01}(a) shows a plot of the surface-surface correlation
function $C^s_L \equiv C(1,L)$ plotted as a function of $1/L$ for $S=1/2$
($\delta=0.3$) and $S=3/2$ ($\delta=0.3$, $0.8$).  In a RS phase one
expects that asymptotically in $L$ the surface-surface correlation
vanishes as $C_L^s \sim L^{- \eta_{\rm RS}^s}$ with $\eta_{\rm RS}^s=1$
\cite{ijr00}. This is indeed observed in the case $S=1/2$. In the $S=3/2$
case instead $C_L^s$ extrapolates to a finite value which indicates the
presence of a surface ordering phenomenon or, in other words, a first
order surface transition (formally $\eta^s=0$).  The same type of behavior
is observed for the $S=3/2$ chain in absence of randomness \cite{fath}.

To obtain the bulk exponent ($\eta$) we consider $C(1, L/2)$ the
correlation function between an edge and a spin in the middle of
the chain, which decays asymptotically as $C(1,L/2+1) \sim L^{-(\eta
+\eta^s)/2}$. Thus each point of reference brings its local scaling dimension,
$\eta/2$ and $\eta^s/2$, respectively.
In order to eliminate the surface exponent contribution we consider
the ratio
\be
{\cal C}_L \equiv \frac{C(1,L/2+1) C(L/2,L)}{C_L^s}
\label{CL}
\ee
which decays asymptotically as ${\cal C}_L \sim L^{-\eta}$. The plot of
$\ln {\cal C}_L$ vs. $\ln L$ is shown in Fig. \ref{FIG01}(b). In the
pure system the bulk exponent is $\eta =1$, while in an RS phase one
expects $\eta_{\rm RS} =2$. The numerical data for $S=3/2$ are quite
consistent with a $1/L$ decay, which seem to be in disagreement with
the existence of a RS phase at weak disorder, but rather support
the same critical behavior as for the pure system.

\begin{figure}[t]
\includegraphics[width=7.8cm]{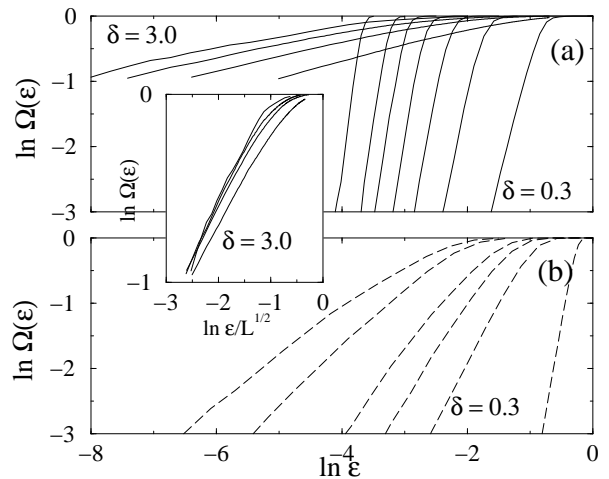}
\caption{Integrated probability distribution of the first gap in log-log
scale.  (a) $S=3/2$, $\delta=0.3$ ($L=4$,$8$,$12$,$16$,$20$,$24$,$32$ from
left to right) and $\delta=3$ ($L=4$,$6$,$8$,$10$) shown as solid lines.
(b) $S=1/2$, $\delta=0.3$ ($L=4$,\ldots,$24$,$32$) shown as dashed lines.
Inset: Collapse of the data for $S=3/2$ and $\delta=3$ as expected in
a RS phase ($L=4,6,8,10$).}
\label{FIG02}
\end{figure}

In the $S=1/2$ case the data are not fully consistent with the expected RS
exponent $\eta_{\rm RS} =2$, shown as a dashed line, but are characterized
by an increasing slope as $L$ increases in the log-log scale, which
is an indication that the asymptotic regime has not yet been reached.
This slope in any case is clearly larger than $1$. Notice that for the
calculation of the bulk exponent we use a combination [see Eq. (\ref{CL})]
of correlation functions between an edge spin and a spin in the middle of
the chain. As the distance between these spins is $L/2$, it is plausible
that ${\cal C}_L$ is plagued by stronger finite size corrections compared
to the surface-surface correlation function $C^s_L$.

\begin{figure}[t]
\includegraphics[width=7.5cm]{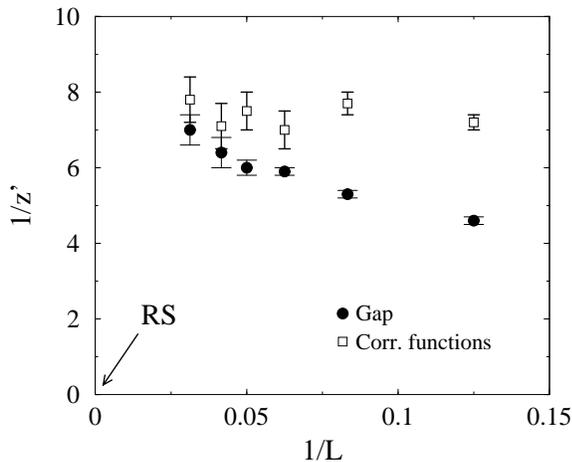}
\caption{Finite size analysis of the inverse dynamical exponent $1/z'$, as
function of the inverse chain length $1/L$ for $S=3/2$ at weak disorder
$\delta=0.3$. In the RS phase one would expect $1/z' \to 0$.}
\label{FIG03}
\end{figure}

\subsection{Dynamical exponent}

We consider next the dynamical singularities in the system,
which are related to the integrated probability distribution,
$\Omega(\varepsilon)$, of the smallest energy gap $\varepsilon$.
This quantity, for small $\varepsilon$, behaves as 
\be
\Omega(\varepsilon) \sim \varepsilon^{1/z'}
\label{omegae}
\ee
If $z'<1$ quantum fluctuations dominate and the true dynamical exponent
is $z=1$, as in the pure system. If $z'>1$ the disorder and quantum
effects compete and the singular properties of the system are governed
by a conventional random fixed point with $z=z'$. In the RS phase $z=z'
\to \infty$.

Figure \ref{FIG02} shows plots of $\ln \Omega(\varepsilon)$ versus $\ln
\varepsilon$ for $S=3/2$ (a) and $S=1/2$ (b). The different scaling
behavior for the two values of the spin can be seen by comparing the
data for the same strength of disorder ($\delta=0.3$).  In the $S=1/2$
chain the distribution becomes broader by increasing the chain length,
while in the $S=3/2$ chain $\ln \Omega(\varepsilon)$ tends to have a
finite non-vanishing slope for small $\varepsilon$, implying that $1/z'>0$,
a conclusion which is at odds with a RS phase.

For sufficiently strong disorder $\delta=3$, however, also the $S=3/2$
spin chains shows a broadening gap distribution (Fig. \ref{FIG02}(a)). The
inset of \ref{FIG02} presents a graph of $\ln \Omega(\varepsilon)$ plotted
as a function of the rescaled variable $\ln \varepsilon/L^\psi$, with
$\psi=1/2$, as expected in the RS phase \cite{nota}. Similar results
were also found at $\delta = 2$.  At such strong values of disorder,
due to difficulties in the convergence of the DMRG method, we restricted
ourselves to exact diagonalization data of rather short chains $L \leq
10$. The asymptotic behavior is however reached already for rather small
chains, as the good collapse of the data of the inset of Fig. \ref{FIG02}
demonstrates. 

We have also calculated the dynamical exponent from the distribution
of the end-to-end correlation functions $C^s_L$, which is expected to
scale as \cite{ijl01}
\be
P_e (C^s_L) \sim \left( C^s_L \right)^{1/z'}
\label{P_e}
\ee
Figure \ref{FIG03} shows a plot of the exponent $1/z'(L)$ vs. $1/L$,
for $\delta=0.3$ obtained from the analysis of the gap and of the
correlation function (Eqs. (\ref{omegae}) and (\ref{P_e})).
Both estimates lead to a finite value $z' \approx 0.13$.
One may wonder whether this is a
genuine effect or just due to a very slow convergence to a possible 
asymptotic behavior consistent with a RS phase $1/z'(L) \to 0$. On the
view of our numerical results we believe that the latter scenario is rather
unlikely, as it would require also a non-monotonic scaling in $L$
of $1/z'(L)$ for the data obtained from the gap.

In a similar way we estimated the dynamical exponent for other strengths
of disorder. For $\delta=0.8$ we find $z'=0.8(2)$. At $\delta=2$ and
even stronger disorder the scaling is consistent with a RS behavior. The
borderline between the RS phase and the weak disorder phase seems to be
located at around $\delta=1$.

\subsection{About the Harris Criterion}

The numerical results for the correlation functions and for the
gap distribution strongly suggest that at weak disorder the $S=3/2$
chain has the same critical properties as in the pure case, i.e. the
disorder is {\it irrelevant} \cite{periodic}.  For sufficiently strong
disorder the DMRG results show also that the system is in a RS phase,
therefore there is a transition induced by the strength of disorder.
The analysis of both the gap and correlation functions indicate that
the region of irrelevance extends up to $\delta=1$.

According to the Harris criterion \cite{harris} disorder is {\it
relevant} if an appropriately defined correlation length \cite{chayes}
in the pure system diverges with an exponent $\nu$ such that $2/\nu>d$
(here $d=1$). In the present case, as the disorder is coupled to the
local energy operator, $\vec{S}_i \cdot \vec{S}_{i+1}$, the correlation
length $\xi$ entering in the Harris criterion is that associated to the
strength of the dimerization, $t$, defined as the amplitude of the alternated
modulation of the coupling constants: $J_i = 1 + (-1)^i t$.
In the limit $t \to 0$ this correlation length diverges as $\xi \sim
|t|^{-\nu}$.  In the $S=1/2$ case, $\nu=2/3 < 2$ \cite{cross_fisher},
implying relevance of weak disorder.

\begin{figure}[t]
\includegraphics[width=7.8cm]{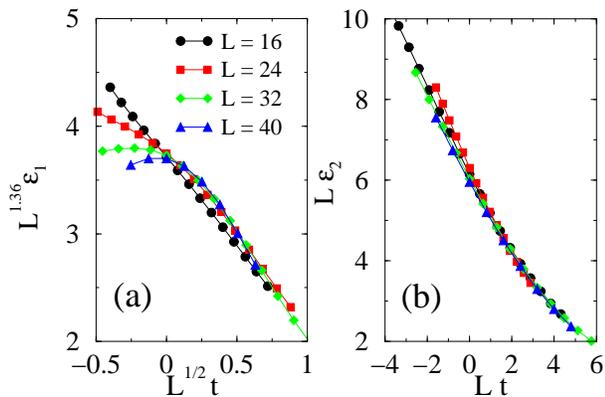}
\caption{Finite size scaling of the first (a) and second (b) gap in the
dimerized $S=3/2$ chain. Data collapse for the first gap is expected to
work only in the disordered region ($t > 0$).}
\label{FIG04}
\end{figure}

\begin{figure}[t]
\includegraphics[width=7.8cm]{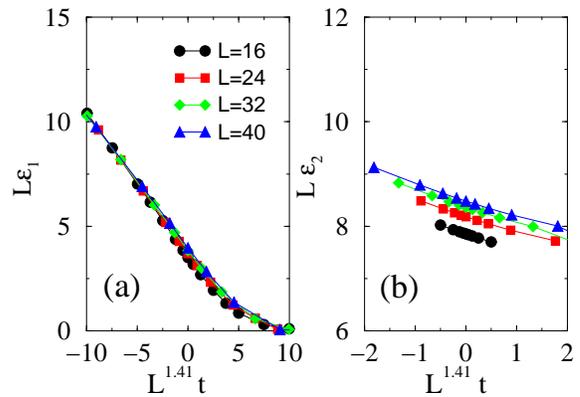}
\caption{As in Fig. \ref{FIG04} for $S=1/2$.}
\label{FIG05}
\end{figure}

In the $S=3/2$ case there is no numerical result available for the
critical exponent, $\nu$, therefore we have analyzed the scaling behavior
of the smallest gaps as function of $t$. Scaling plots are given in
Fig. \ref{FIG04} and as a comparison the same quantities for the $S=1/2$
chain are shown in Fig. \ref{FIG05}.  The gaps are expected to scale as:
\be
\varepsilon_k (L,t) = L^{-z_k}\widetilde{\varepsilon}_k(L^{1/\nu_k}t) ,
\label{gaps}
\ee
with $\widetilde{\varepsilon}_k$ a scaling function and where
$k=1,2$ labels the first and second gaps \cite{gap_scaling}. As
shown above for the $S=3/2$ at weak disorder and for the pure system
the surface is ordered at $t \le 0$, i.e. the correlation function
between the edge spins is finite in the limit $L \to \infty$ (for $L$
even). There are several known examples of models with spontaneous surface
order\cite{igloiturban93}, as the one observed the $S=3/2$ chain. In these
cases one expects a localized low energy excitation at $t=0$ with a gap
that vanishes faster than $1/L$ because the ground state is degenerate
in the thermodynamic limit, i.e.  $\varepsilon_1 \sim L^{-z_s}$ with
$z_s>1$. Indeed this can be seen in Fig. \ref{FIG04} for the smallest
gap in the $S=3/2$ case, $\varepsilon_1$. In absence of dimerization this
quantity scales with an exponent $z_s \approx 1.36 > 1$, as shown by the
scaling collapse at the point $t=0$ in Fig. \ref{FIG04}(a). The figure
also shows that in the surface disordered region $t > 0$ a reasonably good
data collapse for $\varepsilon_1$ is obtained with the choice $L^{1/\nu_s}
t$ with $\nu_s = 2$. Notice that in the surface ordered region $t < 0$
no scaling collapse can be observed. All the other gaps in the $S=3/2$ are
non-localized and scale with $z=1$, as illustrated in Fig. \ref{FIG04}(b)
for $\varepsilon_2$, and their scaling form involves the bulk correlation
length, $\xi \sim |t|^{-\nu}$, with $\nu \approx 1$. This estimate for
$\nu$ is different from the value $\nu=2/3$ in the $S=1/2$ chain. We
expect that the origin of this difference is due to the existence of
a dangerous irrelevant scaling variable \cite{div}, which is generally
observed at first-order surface transitions \cite{igloiturban93} as in
this case.

Our numerical results, summarized above, can be compared with theoretical
predictions in Ref.\cite{Ng}. In the resonating valence bond picture
the low energy excitations of the open $S=3/2$ chain are expected to be
described by effective $S=1/2$ edge-spin degrees of freedom, which are
very weakly coupled to the rest of an $S=1/2$ chain.  In the thermodynamic
limit the edge spins are expected to be decoupled, the effective coupling
being logarithmically small, as $1/\ln L$. In this picture the smallest
gap (at $t=0$) scales as $\varepsilon_1 \sim 1/L \ln L$, whereas the
other gaps scale in the same way as for the open $S=1/2$ chain. Our
numerical findings disagree with this prediction at two points. First,
the localized gap has a faster size-dependence, our numerical results
can not be described by a logarithmic correction. As a matter of fact
for sizes we used in the calculation the effective exponent, $z_s$, is
monotonously increasing with the size and there is no sign of reversing
this tendency. Data of previous numerical calculations in Fig. 5 of
Ref.\cite{QNS} show also strong deviations from a pure logarithmic
correction. Our second disagreement concerns the second gap at $t=0$,
the value of which according to conformal invariance should be:
\be
\varepsilon_2 (L,0)L=\pi v_s x_s
\label{x_s}
\ee
in the large $L$ limit. Here $v_s=3.87$ is the sound velocity
\cite{Hall96} for the $S=3/2$ chain and $x_s$ is the surface anomalous
dimension, which for the $S=1/2$ chain is\cite{ABGR} $x_s=\eta^s/2=1$. For
the $S=3/2$ chain the results in Fig. \ref{FIG04} are considerably
smaller, our estimate is around $x_s=1/2$. Thus there are very probably
new operators for the open $S=3/2$ chain, which are not present in the
$S=1/2$ chain.

As a comparison we plot in Fig. \ref{FIG05} a similar scaling collapse
analysis for the first two gaps in the $S=1/2$ case, using a similar range
of system sizes as in Fig. \ref{FIG04}. Notice that the best collapse
is obtained with a scaling variable $L^{1.41} t$, implying $\nu \approx
0.71$, not far from the expected value $\nu = 2/3$, a difference which
can be imputed to logarithmic corrections.

The analysis at finite $t$ reveals that the dimerization couples
differently to the $S=3/2$ chain with respect to the $S=1/2$ chain.
We argue that the appropriate correlation length to be used in the Harris
criterion \cite{harris,chayes} for the $S=3/2$ chain is $\xi_s$ that
associated to the smallest gap and thus with $\nu_s = 2$ the disorder is
marginally irrelevant.  An accumulation of weak bonds along the chain may
result indeed in an effective cut of the system. The resulting scenario is
that of weakly interacting segments of finite length in which the relevant
length scale is $\xi_s$, that associated to a chain with open boundary
conditions. (We note that a similar scenario is used for the renormalization
of random $S=1$ chains, in which the effective coupling between spin-$1/2$
degrees of freedom is exponentially small\cite{Hyma97}.) This scenario
therefore provides a plausible explanation for
the observed irrelevance of weak disorder for the random $S=3/2$ chains.

\section{Conclusion}

\begin{figure}[t]
\includegraphics[width=6.2cm]{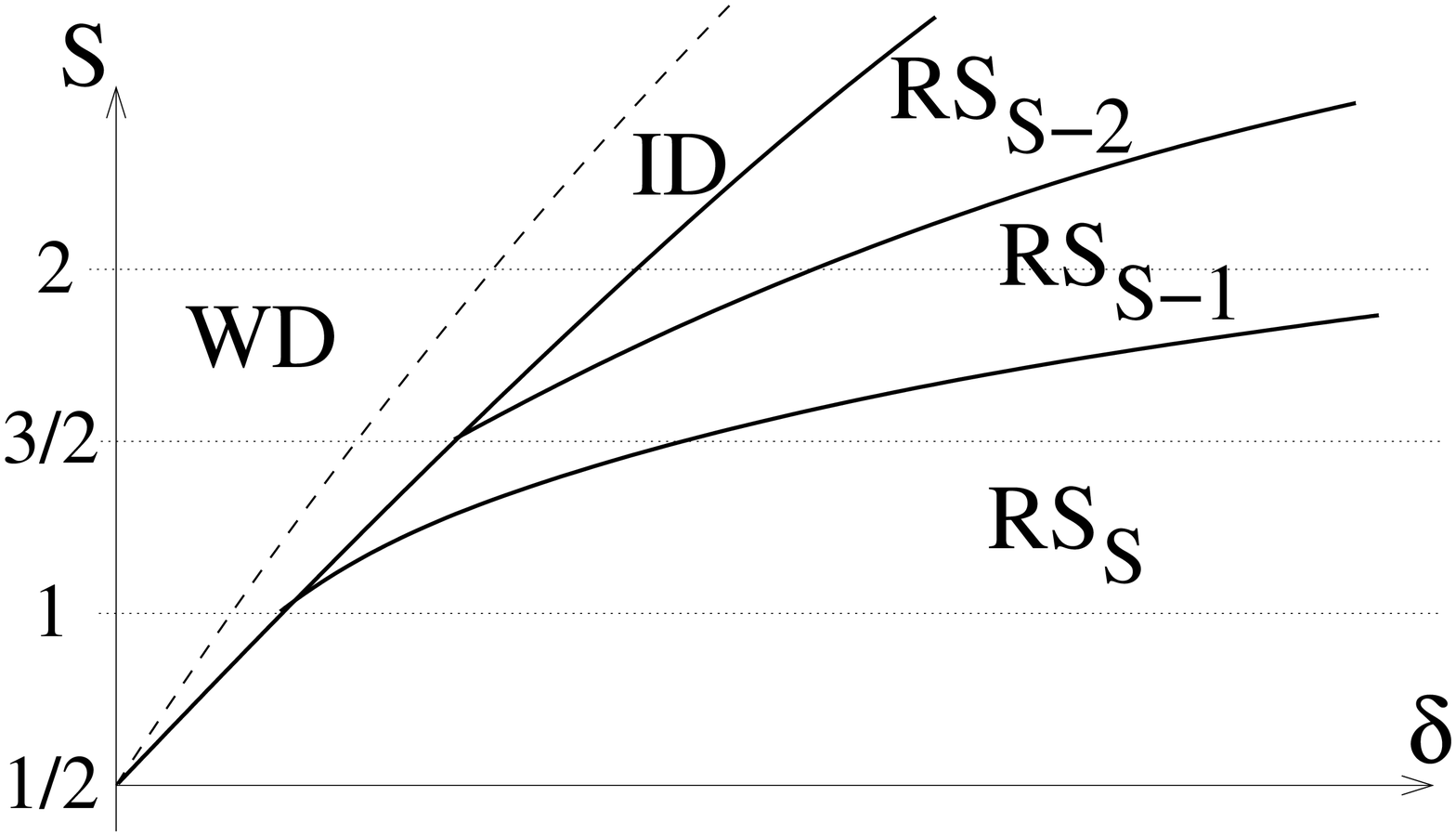}
\caption{Schematic phase diagram of the spin $S$ AFH chain as function of
the strength of disorder $\delta$. The RS$_S$ denotes a random singlet
phase where the relevant degrees of freedom are effective $S$-spins.
The WD and ID are the region at weak and intermediate disorder.}
\label{FIG06}
\end{figure}

We conclude by suggesting a general phase diagram for the spin-S AFH
chains as a function of the strength of disorder $\delta$. This phase
diagram, shown in Fig. \ref{FIG06}, is obtained by combining DMRG and SDRG
results known for the $S=1/2$, $S=1$ and $S=3/2$ cases with some general
arguments for higher $S$.  First of all, differently from the $S=1/2$
case, where any amount of disorder is known to lead to a RS phase,
we expect that all chains for $S > 1/2$ have a region of irrelevance,
the weak disorder (WD) region, where the system is either critical
(half-integer $S$) or gapped (integer $S$). This was already known for
$S$ integer. Our results for the $S=3/2$ chain suggest that it may be
true also for non-integer spin.

At strong enough disorder all $S$ are in a RS phase composed of
effective $S$ spins, which we indicate as RS$_{\rm S}$.  For the $S=1$
case there is only one such phase (RS$_{\rm 1}$), while a recent
SDRG study for $S=3/2$ \cite{Refa02} predicts two distinct phases:
RS$_{\rm 3/2}$ and RS$_{\rm 1/2}$.  We found no signatures of this
spin reduction transition in our DMRG calculations, which could be
anyhow difficult to detect as the RS$_{\rm 1/2}$ and RS$_{\rm 3/2}$
have identical critical exponents.  There is also the possibility to
have, for higher spins ($S > 3/2$), a sequence of multiple RS phases
of different nature. Possible phase diagram for higher $S$ have been
recently discussed \cite{Daml02}.  We have indicated these transition
lines in the phase diagram of Fig. \ref{FIG06}, which, for the time
being cannot be supported by numerical results.  Finally, the WD region
may be separated from the RS phase(s) by an intermediate disorder (ID)
region where exponents vary continuously with $\delta$, as observed in
other models \cite{Carl01}. For the gapped (integer S) case this would
correspond to a region of Griffiths singularities. The qualitative nature
of the disorder induced cross-over phenomena for the $S=1$ chain is still
debated (see in Refs.\cite{Hida99,qmc}). We will discuss this issue in
a future publication \cite{lcri}.

F.I. is grateful to G. F\'ath for useful discussions. This work has been
supported by a German-Hungarian exchange program (DAAD-M\"OB), by the
Hungarian National Research Fund under grant No OTKA TO34138, TO37323,
MO45596 and M36803, by the Ministry of Education under grant No. FKFP
87/2001, by Kuwait University Research Grant No. [SP 09/02]. by the
Centre of Excellence ICA1-CT-2000-70029,  and the numerical calculations
by NIIF 1030.  P.L. thanks M.M. Sharma for help in numerical calculation.

{\it Note added} -- After this paper was submitted we became aware of
a recent work of Seguia {\it et al.} \cite{segu03}, who performed a
SDRG analysis of the disordered $S=3/2$ AFH chain. Their calculation
differs from that of Ref. \cite{Refa02} as the decimation scheme keeps
into account more states, thus it is expected to be more accurate.
While the results of Ref. \cite{Refa02} indicated the existence of
two RS phases with different effective spins ($S_{\rm eff} = 1/2$ and
$S_{\rm eff} = 3/2$), in the work of Seguia {\it et al.} \cite{segu03}
the renormalization flow indicates that the weak disorder is an irrelevant
perturbation of the system. This is in agreement with the conclusion of
this paper.


\begin{thebibliography}{}

\bibitem{Hald81} F. D. M. Haldane, \prl {\bf 50}, 1153 (1983).

\bibitem{s32} 
        H. J. Schulz, \prb {\bf 34}, 6372 (1986); 
        I. Affleck and F. D. M. Haldane, \prb {\bf 36}, 5291 (1987); 
        I. Affleck {\it et al.} J. Phys. A {\bf 22}, 511 (1989).

\bibitem{Hall96}
        K. Hallberg {\it et al.} \prl {\bf 76}, 4955 (1996).

\bibitem{denN89} 
        M. den Nijs and K. Rommelse, \prb {\bf 40}, 4709 (1989).

\bibitem{mdh}
        S. K. Ma, C. Dasgupta and C.-K. Hu, Phys. Rev. Lett. {\bf 43}, 1434 (1979);
        C. Dasgupta and S. K. Ma, Phys. Rev. B{\bf 22}, 1305 (1980).

\bibitem{fisherxx}
        D. S. Fisher, Phys. Rev. B {\bf 50}, 3799 (1994).

\bibitem{henelius}
        P. Henelius and S. M. Girvin, Phys. Rev. B {\bf 57}, 11457 (1998).

\bibitem{ijr00}
        F. Igl\'oi, R. Juh\'asz, and H. Rieger, Phys. Rev. B {\bf 61},
        11552 (2000).

\bibitem{stolze}
        K. Hamacher, J. Stolze, and W. Wenzel, 
	Phys. Rev. Lett. {\bf 89}, 127202 (2002);
        N. Laflorencie and H. Rieger, 
	Phys. Rev. Lett. {\bf 91}, 229701 (2003)

\bibitem{griffiths}
        R. B. Griffiths, Phys. Rev. Lett. {\bf 23}, 17 (1969).

\bibitem{Hyma97} 
        R. A. Hyman and K. Yang, \prl {\bf 78}, 1783 (1997);
        C. Monthus, O. Golinelli and T. Jolicoeur, \prl {\bf 79},3254 (1997);
        A. Saguia, B. Boechat, and M.~A.~Continentino, 
        Phys. Rev. Lett. {\bf 89}, 117202 (2002).

\bibitem{Refa02} 
        G. Refael, S. Kehrein, and D. S. Fisher, \prb {\bf 66}, 
        R060402 (2002).

\bibitem{Daml02} 
        K. Damle, \prb {\bf 66}, 104425 (2002); 
        K. Damle and D. A. Huse, \prl {\bf 89}, 277203 (2002).

\bibitem{DMRGbook}
        {\it Density Matrix Renormalization: A New Numerical
        Method in Physics}, edited by I. Peschel, X. Wang, M. Kaulke,
        and K. Hallberg (Springer, Berlin, 1999).

\bibitem{fath}
        G. F\'ath and F. Igl\'oi, unpublished.

\bibitem{nota} 
        For strong disorder we restricted ourselves to exact 
        diagonalization as DMRG is not reliable in that regime.

\bibitem{ijl01}
        F. Igl\'oi, R. Juh\'asz and P. Lajk\'o, \prl {\bf 86}, 1343 (2001).

\bibitem{periodic}
        The irrelevance of disorder is not related to the use of open boundaries as we checked for the gap distribution on small periodic chains.

\bibitem{harris}
        A.B. Harris, J. Phys. C{\bf 7}, 1671 (1974).

\bibitem{chayes}
        J.T. Chayes {\it et al.} \prl {\bf 57}, 2999 (1986).

\bibitem{cross_fisher}
        M. C. Cross and D. S. Fisher, Phys. Rev. B {\bf 19}, 402 (1979);
        T. Papenbrock {\it et al}, cond-mat/0212254.

\bibitem{gap_scaling} In Eq. (\ref{gaps}) we leave the possibility of having
        different gaps scaling with different exponents (i.e. $z$ and $\nu$
        may depend on $k$). This is indeed found in the $S=3/2$ chain. 

\bibitem{igloiturban93}
        F. Igl\'oi and L. Turban, Phys. Rev. B{\bf 47}, 3404 (1993).

\bibitem{div}
        M. E. Fisher, in {\it Renormalization Group in Critical Phenomena
        and Quantum Field Theory}, proceedings of a conference, edited by
        J. D. Gunton and M. S. Green (Temple University, Philadelphia)(1974).

\bibitem{Carl01}
        E. Carlon, P. Lajk\'o, and F. Igl\'oi, \prl {\bf 87}, 277201 (2001).

\bibitem{Ng}
        T-K. Ng, \prb {\bf 50}, 555 (1994).

\bibitem{QNS}
        S. Qin, T-K. Ng, and Z-B. Su, \prb {\bf 52}, 12844 (1995).

\bibitem{ABGR}
        F.C. Alcaraz, M. Baake, U. Grimm, and V. Rittenberg, J. Phys. A{\bf 21}, L117 (1988).

\bibitem{Hida99}
        K. Hida, \prl {\bf 83}, 3297 (1999); 
        K. Yang and R. A. Hyman, \prl {\bf 84}, 2044 (2000); 
        K. Hida, \prl {\bf 84}, 2045 (2000).

\bibitem{qmc}
        S. Todo, K. Kato, and H. Takayama, J. Phys. Soc. Jpn. {\bf 69}, 
        355 (2000);
        S. Bergkvist, P. Henelius, and A. Rosengren, \prb {\bf 66},
        134407 (2002). 

\bibitem{lcri}
        P. Lajk\'o, E. Carlon, H. Rieger, and F. Igl\'oi (unpublished).

\bibitem{segu03} 
        A. Saguia,  B. Boechat, and M. A. Continentino, \prb {\bf 68} 020403(R) (2003).

\end{thebibliography}
\end{document}